\documentclass[conference,twocolum,final]{IEEEtran}

\usepackage{bbm}
\usepackage{algorithmic}
\usepackage{algorithm} 
\usepackage{amsthm}
\usepackage{amsmath,amssymb}
\usepackage{subfigure}
\usepackage{graphics}
\usepackage{epstopdf}
\DeclareGraphicsExtensions{.eps}

\ifCLASSINFOpdf 
\usepackage[pdftex]{graphicx} 
\usepackage[pdftex,colorlinks,
           draft, 
           bookmarks=true,%
            pdftitle=Optimal\ Geographic\ Caching\ In\ Cellular Networks,%
           pdfauthor=B.\ Blaszczyszyn\ -\ A.\ Giovanidis]{hyperref}  
\usepackage[numbers,sort&compress]{natbib} 
\else

\usepackage[dvips]{graphicx} 
\usepackage[dvips,colorlinks, draft, 
           bookmarks=true,%
           pdftitle=Optimal\ Geographic\ Caching\ In\ Cellular Networks,%
           pdfauthor=B.\ Blaszczyszyn\ -\ A.\ Giovanidis]{hyperref}  
 \usepackage[numbers,sort&compress]{natbib} 
\fi 
\hypersetup{
   bookmarksnumbered,
   pdfstartview={FitH},
   citecolor={blue},
   linkcolor={red},
   urlcolor=[rgb]{0,0.55,0},
   pdfpagemode={UseOutlines}}

\newtheorem{Pro}{Proposition}

\newtheorem{Lem}{Lemma}

\newtheorem*{proof*}{Proof}

\begin{document}

\title{Coverage Gains from the Static Cooperation of Mutually Nearest Neighbours}

\author{Luis David \'Alvarez-Corrales$^\star$, Anastasios~Giovanidis$^\star$ and Philippe Martins$^\star$\\[2ex]}

\maketitle

\begin{abstract}
Cooperation in cellular networks has been recently suggested as a promising scheme to improve system performance. In this work, clusters are formed based on the Mutually Nearest Neighbour relation, which defines which stations cooperate in pair and which do not. When node positions follow a Poisson Point Process (PPP) the performance of the original clustering model can be approximated by another one, formed by the superposition of two PPPs (one for the singles and one for the pairs) equipped with adequate marks. This allows to derive exact expressions for the network coverage probability under two user-cluster association rules. Numerical evaluation shows coverage gains from different signal cooperation schemes that can reach up to 15\% compared to the standard non-cooperative network coverage. The analysis is general and can be applied to any type of cooperation or coordination between pairs of transmitting nodes.
\end{abstract}


\let\thefootnote\relax\footnotetext{\hspace{-2ex}$^\star$ Universit\'e Paris-Saclay, T\'el\'ecom ParisTech \& CNRS-LTCI, 23 avenue d'Italie, 75013, Paris, France. \\ 
emails: \{luis.alvarez-corrales, anastasios.giovanidis, philippe.martins\} @telecom-paristech.fr }
\newcommand{\thefootnote}{\arabic{footnote}}

\section{Introduction}

This work investigates the potential performance gains that can be obtained in a cellular network, by coordination or cooperative transmission between base stations (BSs). Cooperation is particularly relevant for users located at the cell-edge, where significant $\mathrm{SINR}$ gains can be achieved in the downlink. In the wireless literature there is a considerable amount of research on the topic, which relates to the concept of CoMP, Network MIMO \cite{GesMultMIMO2010}, 
\cite{GioA012012} (see also references therein), or C-RAN. The various strategies proposed differ in the number of cooperating nodes, the type of signal cooperation, the amount of information exchange, and the way groups (clusters) are formed either dynamically or in a static way a-priori chosen.

It is possible to analyse such cooperative networks by use of stochastic geometry. Application of point processes to model the positions of wireless nodes gives the possibility to include the impact of irregularity of BS locations on the users' performance (e.g. $\mathrm{SINR}$, throughput, delay). Furthermore, the gains from cooperation can be quantified in a systematic way, so that one need not test each different instance of network topology by simulations. Closed formulas are very important for an operator that wants to plan and deploy an infrastructure with cooperation functionality, because these can provide intuition on the relative influence of various design parameters.

There are considerable results available in this area. In \cite{BacAStoGeo2015}, Baccelli and Giovanidis analyse the case where BSs are modelled by a Poisson Point Process (PPP) and each user-terminal triggers the cooperation of its two closest BSs for its service. The authors show coverage improvements and an increase of the coverage cell. The work is extended in \cite{NigCoordMul2014} by Nigam et al for larger size of clusters, showing that BS cooperation is more beneficial for the worst-case user. The $\mathrm{SINR}$ experienced by a typical user when served by the $K$ strongest BSs is also investigated by Blaszczyszyn and Keeler in \cite{BlasStuSINTFact2015}, and the authors derive tractable integral expressions of the coverage probability for general fading by use of factorial moment measures. An analysis of a similar problem with use of Laplace transforms (LT) is provided by Tanbourgi et al in \cite{TanTracMod2014}.

All the above works assume that a user-terminal \textit{dynamically} selects the set of stations that cooperate for its service. Such an assumption is difficult to be applied in practice because the cluster formation should change with every different configuration of users. For this reason we propose here to group BSs \textit{in a static way}, so that clusters are a-priori defined and do not change over time. The appropriate static clustering should result in considerable performance benefits for the users, with a cost-effective infrastructure. In favour of the static grouping approach are other authors as well, like Akoum and Heath \cite{AkouIntCoor2013} who randomly group BSs around virtual centres, and Guo et al who analyse in \cite{GuoSPGP2014} the coverage benefits of cooperation pairs modelled by a Gauss-Poisson point process.

The existing static clustering models are not sufficient. They either group BSs in a random way \cite{AkouIntCoor2013}, or they randomly generate additional cluster nodes around a cluster center \cite{GuoSPGP2014,AfshFundClustCent}, which is translated in the physical world into installing randomly new nodes in the existing infrastructure. A more appropriate analysis should have a map of existing BS locations as the starting point, and from this define in a systematic way cooperation groups. The criterion for grouping should not be random, but rather node proximity, in order to limit the negative influence of first-order interference. For these reasons the authors propose in \cite{GioAnalInt2015} a grouping method based on a variation of the Nearest Neighbour model \cite{HaggNNHS1996}. The analysis was restricted to the case where only singles and pairs are allowed. Having a PPP as a basis for the BS locations, pairs are selected as the nodes that satisfy the \textit{mutually nearest neighbour relation}, whereas the process of singles is formed by the remaining nodes. The analysis provided structural characteristics and expressions for the expectation and LT of the interference generated by each one of the two resulting processes.

The processes of singles and pairs were shown not to be PPPs, so the derivation of numerically tractable formulas for $\mathrm{SINR}$ related metrics, as done in \cite{AndATract2011}, is not evident. To find such formulas for the Nearest Neighbour cooperation we introduce, in Section II, an approximative model: the superposition of two independent PPPs, equipped with structural characteristics of the original single and pair processes. Potential signals emitted by the cooperative nodes, and their generalisation, are proposed in Section III. Section IV analyses the interference in the superposition model. Analysis of the coverage probability for two scenarios of user-to-BS association is provided in Section V. The analytical formulas are validated through simulations and the gains of static nearest neighbour grouping are quantified. The final conclusions are drawn in Section VI. \textit{Proofs and supplementary material can be found in the arXiv version of the work, with the same title}. 

Note that our approach is general. It can be applied to many cooperation variations, ranging from simple coordination of the BSs in group, to fully cooperative transmission using knowledge of the channel states.

\section{Model under study}\label{secII}
On $\mathbb{R}^2$ we model the BS locations by a stationary PPP $\Phi$ with density $\lambda>0$. We assume that a user-terminal is located at the Cartesian origin $\left(0,0\right)$ and we examine the network performance at its position. This is the \textit{typical user approach}. We denote by $\|\cdot \|$ the Euclidean distance in $\mathbb{R}^2$.

\subsection{Previous results on the topic}

\begin{figure}[htbp]
\centering
\subfigure{\includegraphics[width=0.3\textwidth]{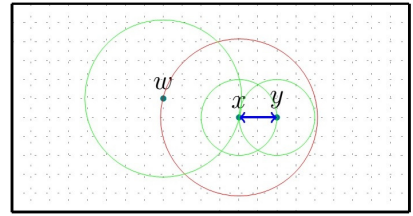} \label{SinAndPairs1}}
\subfigure{\includegraphics[clip,width=0.37\textwidth]{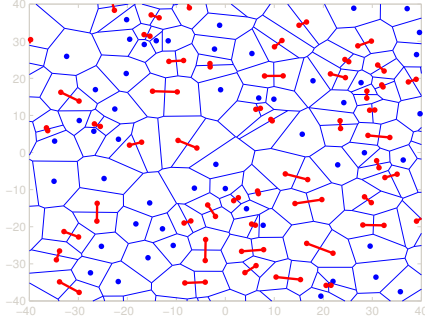}\label{SinAndPairs2}}
\caption{ \textit{Upper image}: since the $x$ and $y$ atoms are mutually nearest neighbours, they work in pair. $x$ is the closest atom of $w$, but $w$ is not the closest atom to $x$, thus, $w$ is a single. \textit{Lower image}: a Poisson realisation. The blue dots are the single BSs, the red dots are the cooperating pairs. \label{SinAndPairs}}
\end{figure}

In \cite{GioAnalInt2015}, starting from some given point process (e.g. PPP) the authors propose a grouping method of BSs based on the Nearest Neighbour model \cite{HaggNNHS1996}. Two BSs belong to the same cooperating pair if each one of them is the nearest neighbour of the other \cite[Def. 1]{GioAnalInt2015}. If a BS does not form a mutually Nearest Neighbour pair with any other BS from the process, it remains single \cite[Def. 2]{GioAnalInt2015}. This is explained graphically in Figure \ref{SinAndPairs}. In this way, $\Phi$ splits into a process of single points and another one of the pairs. Figure \ref{SinAndPairs} also displays an example of a Poisson realisation, showing the singles and the pairs that can be formed. Let $\gamma:=\frac{2}{3}-\frac{\sqrt{3}}{2\pi}$ and $\delta:=\frac{1}{2-\gamma}\approx 0.6215$. It has been proven that, in accordance with the number $\delta$, in average $62.15\%$ of BSs are in pair and $37.85\%$ of BSs are singles \cite[Cor. 1]{GioAnalInt2015}. As a result, the processes of singles and pairs are not Poisson. The authors also have shown that the distance between cooperating BSs follows a Rayleigh distribution, with scale parameter $\alpha:=(2\lambda\pi(2-\gamma))^{-1/2}$ \cite[Th. 2]{GioAnalInt2015}. From now on, we refer to the previous model as the \textbf{Nearest Neighbour (NN) model}. 

As a consequence of the non-Poissonian behaviour, it is difficult to make a complete performance analysis for SINR related metrics. Instead, we use in this work the following model to approximate these metrics.

\subsection{Poisson superposition model}

To imitate the process of singles, we consider a PPP $\hat{\Phi}^{(1)}$, with parameter $(1-\delta)\lambda$. In this way, the new process for singles has the same expected value with the one from the NN model \cite[Cor. 1]{GioAnalInt2015}.

To imitate the process of pairs, we also consider a PPP $\hat{\Phi}^{(2)}$, independent of $\hat{\Phi}^{(1)}$, with intensity $\frac{\delta}{2}\lambda$. We call the atoms of this process \textit{the parents}. The process $\hat{\Phi}^{(2)}$ is independently marked. Each mark of a parent represents its pairing BS, \textit{the daughter}. The idea is that each couple $(\textit{parent},\textit{daughter})$ imitates a cooperating pair of the NN model. Let us consider $(Z_r)_{r>0}$ a family of independent random variables, independent also of $\hat{\Phi}^{(1)}$ and of $\hat{\Phi}^{(2)}$, where each $Z_r$ follows a Rice distribution, with parameter $(r,\alpha)$. If $Y$ is a random point representing a parent, then we define its mark by $Z_{\|Y\|}$.

\begin{figure}
\centering
\includegraphics[width=0.2\textwidth]{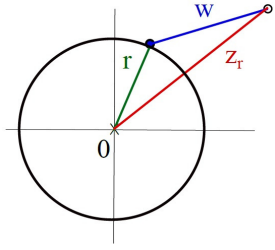}
\caption{Two cooperating BSs, where $r$ and $Z_r$ are their distances from the origin, and $W$ is the distance between them. \label{Zr}}
\end{figure}

To understand the choice for the marks, suppose that a BS is placed at the polar coordinates $(r,\theta)$, with $r>0$ and $\theta\in [0,2\pi)$ fixed (see Figure \ref{Zr}). Assume also that this BS belongs to a cooperating pair from the NN model, and let us denote by $W$ the distance between the stations in pair, which is Rayleigh distributed \cite[Th. 2]{GioAnalInt2015}, with scale parameter $\alpha$. If $Z$ denotes the distance from the typical user to the second BS, the isotropy of the PPP implies that the distribution of $Z$ is independent of $\theta$. Moreover, we have the following result.

\begin{Pro}\label{riceProp}
The random variable $Z$ is Rice distributed, with parameters $(r,\alpha)$. The probability density function (PDF) of $Z$ is given by 
\begin{equation}\label{riceDen}
f(z|r)=\frac{z}{\alpha^2}e^{-\frac{z^2+r^2}{2\alpha^2}} I_0\left( \frac{zr}{\alpha^2} \right),
\end{equation}
where $I_0(x)$ is the modified Bessel function, of the first kind, with order zero. 
\end{Pro}
The angular coordinate of a PPP atom is uniformly distributed in $[0,2\pi)$. Moreover, the Cartesian coordinates of a point around a center, with Rayleigh radial distance and uniform angle, are distributed as an independent Gaussian vector \cite[pp. 276, Ex. 7b]{AfirstCourseRoss}. Given this, Proposition \ref{riceProp} follows from \cite{AfshFundClustCent}.   

Other models where the angle plays a role can be treated, as well, by adding further marks in the $\hat{\Phi}^{(2)}$ process. 

\subsection{The distribution of the closest distances} \label{nearestN}

Let $R_1$ and $R_ 2$ denote the random variables of the distances from the closest element of $\hat{\Phi}^{(1)}$ and $\hat{\Phi}^{(2)}$ to the origin, respectively. Denote also by $Z_2$ the mark of the parent at $R_2$. It is known that the random variables $R_1$ and $R_2$ are Rayleigh distributed \cite{AndATract2011}, with scale parameters $\xi$ and $\zeta$, where $\xi:=((1-\delta)2\lambda \pi)^{-1/2}$ and $\zeta:=(\delta \lambda \pi)^{-1/2}$. By definition, $R_2$ and $Z_2$ are not mutually independent, but we can derive their joint PDF.

\begin{Lem}\label{denR2ZR2}
The joint PDF of the random variable $(R_2,Z_2)$ is given by 
\begin{equation} \label{jointDensityFunction}
\begin{split}
f(r,z) & = \frac{rz}{(\alpha \zeta)^2} e^{-\frac{r^2}{2}\left(\frac{1}{\alpha^2}+\frac{1}{\zeta^2} \right)-\frac{z^2}{2\alpha^2}}I_0\left( \frac{rz}{\alpha^2} \right).
\end{split} 
\end{equation}
Furthermore, the random variable $Z_2$ is Rayleigh distributed, with scale parameter $( \alpha^2+\zeta^2)^{1/2}$.
\end{Lem}

In section \ref{secV} we make use of the distribution of the random vector $(R_1,R_2,Z_2)$. Because $R_1$ is independent of $(R_2,Z_2)$, the joint PDF is the product of the PDF of $R_1$ with the joint PDF of $(R_2,Z_2)$.

\section{Received Signals}\label{secIV}

In this work we restrict ourselves to omnidirectional BSs. This allows for a class of signals which is large enough. The directional case can also be included, but it requires extra integration with respect to angles, which unnecessarily complicates the analysis without substantial difference \cite[Section V]{BacAStoGeo2015}. 
 
We consider an independent, identically distributed family $(h_r)_{r>0}$ of positive exponential variables, with parameter $1$, also independent of the other random elements introduced in Section \ref{secII}. With fixed $p>0$, the couple $(h_r,p)$ represents the random propagation effects and the power signal emitted to the typical user from a BS at distance $r>0$ from the origin. Let us also choose the \textit{path-loss} function as $l(r):=\frac{1}{r^\beta}$,  with \textit{path loss exponent} $\beta>2$. 

\subsection{Single atoms}
Consider a generic random field $f:[0,\infty) \longrightarrow \mathbb{R}^+$, independent of other random objects. The quantity $f(r)$ represents the received signal at the typical user, when transmitted by a single BS, whose distance to the origin is $r>0$. For a single BS, whose distance to the origin is $r>0$, it is natural to consider 
\begin{equation}\label{signalsin}
f(r)=p \frac{h_r} {r^\beta},
\end{equation}  
which follows an exponential distribution, with parameter $\frac{r^\beta}{p}$. 

\subsection{Pair cooperation}
To keep cooperation general, consider a generic random field $g:[0,\infty)\times [0,\infty) \longrightarrow \mathbb{R}^+$, independent of other random objects, where the quantity $g(r,z)$ represents the received signal at the typical user, when transmitted by a pair of BSs whose distances to the origin are $r>0$ and $z>0$, respectively. The received signal can take the following example expressions, which refer to different types of cooperation or coordination,
\begin{equation}\label{coopFunction}
g(r,z)=
\begin{cases}
p \frac{h_r}{r^\beta} + p \frac{h_z}{z^\beta}  \ , \ \ \ \ \ \ \ \ \ \ \ \ \ \ \ \ \ \ \ \ \mbox{\textbf{[NSC]}} \\
\mathbf{1}_{on_r}p \frac{h_r}{r^\beta} + (1-\mathbf{1}_{on_r})p \frac{h_z}{z^\beta} \ , \ \ \mbox{\textbf{[OFF]}} \\
\max \left\{p \frac{h_r}{r^\beta},p \frac{h_z}{z^\beta}\right\} \ , \ \ \ \ \ \ \ \ \ \ \ \ \ \mbox{\textbf{[MAX]}} \\
\left| \sqrt{p \frac{h_r}{r^\beta}}e^{i\theta_r}+\sqrt{p \frac{h_z}{z^\beta}}e^{i\theta_z}\right|^2 \ \ \ \ \mbox{\textbf{[PH]}}
\end{cases}.
\end{equation}
In the above, $(\mathbf{1}_{on_r})_{r>0}$ and $(\theta_r)_{r>0}$ are two different families of indexed identically distributed random variables, independent of other random objects. They follow a Bernoulli distribution, with parameter $q\in (0,1)$ ($\overline{q}:=1-q$), and a general distribution over $[0,2\pi)$, respectively. \textbf{[NSC]} refers to non-coherent joint transmission, as in  \cite{NigCoordMul2014,BlasStuSINTFact2015,TanTracMod2014,GuoSPGP2014}, where each of the two BSs transmit an orthogonal signal, and the two are added at the receiver side. \textbf{[OFF]} refers to the case where one of the two BSs is active and the other inactive, according to an independent Bernoulli experiment, independent of the BSs positions. \textbf{[MAX]} refers to the case where the BS with the strongest signal is actively serving a user, while the other is off. The \textbf{[OFF]} and \textbf{[MAX]} cases are relevant to energy saving operation. In the \textbf{[PH]} case, two complex signals are combined in phase (see \cite{BacAStoGeo2015,NigCoordMul2014}), in particular, when $cos(\theta_r-\theta_z)=1$, the two signals are in the same direction, and they add up coherently at the receiver (user side), giving the maximum cooperating signal. 

The above expressions in \eqref{coopFunction} are merely examples for the cooperation signals. A more general family can be proposed with specific properties. Consider $c_i:[0,\infty)\times[0,\infty) \longrightarrow \mathbb{R}$, and $d_i:[0,\infty)\times[0,\infty) \longrightarrow \mathbb{R}^+$, for $1\leq i \leq n$, some deterministic functions, and suppose that  
\begin{equation}\label{TAIL}
\mathbb{P}(g(r,z)>T)=\sum^n_{i=1}c_i(r,z)e^{-d_i(r,z)T}.
\end{equation}
When analysing performance related to coverage probability, the tail probability distribution functions (CCDF) for the signals, that can be written as \eqref{TAIL} lead easier to numerically tractable formulas. However, the function defined in \eqref{TAIL} is not necessarily a CCDF. For this to hold, some conditions on the functions $c_1(r,z),\ldots,c_n(r,z),d_1(r,z),\ldots,d_n(r,z)$ should be imposed. Interestingly, the CCDF of $g(r,z)$ in the \textbf{[NSC]}, \textbf{[OFF]}, and \textbf{[MAX]} cases fulfils equation \eqref{TAIL} (see Table \ref{table}). Furthermore, there exist important families of random variables whose CCDF actually has the form described in equation \eqref{TAIL}: the \textit{hypo-exponential distribution}, the \textit{hyper-exponential distribution}, the maximum over a finite number of exponential random variables, etc. 
\begin{table} 
	\centering
	\caption{Expressions for the CCDF and the LT}\label{table}
    \begin{tabular}{ | l | p{4cm} | p{2.5cm} | } 
    \hline
     \ \ & \centering $\mathbb{P}(g(r,z)>T)$  & $\mathbb{E}[e^{-s g(r,z)}]$  \\ \hline
    \textbf{[NSC]} & $\frac{r^{\beta}z^\beta}{p(r^\beta-z^\beta)} \Big(e^{-\frac{z^\beta}{p}T}-e^{-\frac{r^\beta}{p}T} \Big)$ & $\frac{r^\beta}{sp+r^\beta}\frac{z^\beta}{sp+z^\beta}$
    \\ \hline
    \textbf{[OFF]} & $qe^{-\frac{r^\beta}{p}T}+\overline{q}e^{-\frac{z^\beta}{p}T}$ & $q\frac{r^\beta}{sp+r^\beta}+\overline{q}\frac{z^\beta}{sp+z^\beta}$ 
    \\ \hline
    \textbf{[MAX]} & $e^{-\frac{r^\beta}{p}T}+e^{-\frac{z^\beta}{p}T}-e^{- \left( \frac{ r^\beta}{p} + \frac{z^\beta}{p}\right)T}$ & $\frac{r^\beta}{sp+r^\beta}+\frac{z^\beta}{sp+z^\beta}-\frac{r^\beta+z^\beta}{sp+r^\beta+z^\beta}$
    \\    \hline
    \end{tabular}
\end{table}
    

\section{Interference Field}\label{secIII}

For some $r>0$, let us denote by
\begin{subequations}\label{LPfunctions}
\begin{align}
& \mathcal{L}_f(s;r) = \mathbb{E}\left[ e^{-s f(r)}\right], \label{LPfunctionsa} \\
& \mathcal{L}_g(s;r,\rho) = \mathbb{E}\left[ e^{-s  g(r,Z_r)} \mathbf{1}_{\{Z_r>\rho\}}\right], \label{LPfunctionsb}
\end{align}
\end{subequations}
the LT of the signal generated by a single BS and the LT of the signal generated by a cooperation pair, given that the radius of the daughter is larger than $\rho\geq 0$. When $\rho=0$, $\mathcal{L}_g(s;r,0)$ will be denoted just by $\mathcal{L}_g(s;r)$. For example, if we take $f(r)$ as in equation \eqref{signalsin}, we get  
\begin{equation}\label{LPsingles}
\mathcal{L}_f(s;r) = \frac{r^\beta}{sp+r^\beta }. 
\end{equation}
Recall that $p \frac{h_r} {r^\beta}$ and $p \frac{h_z} {z^\beta}$ are independent, exponential random variables, with parameter $\frac{r^\beta}{p}$ and $\frac{z^\beta}{p}$. In Table \ref{table} we find expressions for $\mathbb{E}[e^{-sg(r,z)}]$ in the \textbf{[NSC]}, \textbf{[OFF]}, and \textbf{[MAX]} cases. By remarking that 
\begin{equation*}
\mathcal{L}_g(s;r) = \mathbb{E}\left[\mathbb{E}\left[ e^{-sg(r,Z_r)}\mathbf{1}_{\{Z_r>\rho\}} \Big|Z_r\right]\right],
\end{equation*}
we get analytical expressions for $\mathcal{L}_g(s;r)$ in the \textbf{[NSC]}, \textbf{[OFF]}, and \textbf{[MAX]}. For example, in the \textbf{[NSC]} we have that
\begin{equation*}
\begin{split}
\mathcal{L}_{g}(s;r,\rho) 
& = \frac{r^\beta}{sp+r^\beta} \int^\infty_\rho \frac{z^\beta}{sp+z^\beta} f(z|r)dr,
\end{split}
\end{equation*}
where $f(z|r)$ is the density function of the Rice random variable $Z_r$ (see equation \eqref{riceDen}). For the more general distribution descrived by equation \eqref{TAIL}, it is also possible to give analytical formulas similar to the previous equation. In the \textbf{[PH]} case the expression for $\mathcal{L}_{g}(s;r,\rho) $ is more complicated (see \cite{BacAStoGeo2015} for $cos(\theta_r-\theta_z)=1$). \\

We consider the interference fields generated by all the elements of $\hat{\Phi}^{(1)}$ and $\hat{\Phi}^{(2)}$ outside the radius $\rho$
\begin{subequations} \label{interference}
\begin{align}
\hat{\mathcal{I}}^{(1)}(\rho) & = \sum_{x \in \hat{\Phi}^{(1)}, \|x\|>\rho} f(\|x\|), \label{interference2a} \\
\hat{\mathcal{I}}^{(2)}(\rho) & = \sum_{\stackrel{y \in \hat{\Phi}^{(2)}}{ \|y\|>\rho,Z_{\|y\|}>\rho}} g(\|y\|,Z_{\|y\|}). \label{interference2b}
\end{align}
\end{subequations}
When $\rho=0$, they are just denoted by $\hat{\mathcal{I}}^{(1)}$ and $\hat{\mathcal{I}}^{(2)}$. The total interference generated outside possibly different radii for the two processes, i.e. $\rho_1>0$ and $\rho_2>0$ is
\begin{equation}\label{totalInt}
\hat{\mathcal{I}}(\rho_1,\rho_2):=\hat{\mathcal{I}}^{(1)}(\rho_1)+\hat{\mathcal{I}}^{(2)}(\rho_2).
\end{equation}
When $\rho_1=\rho_1=0$, we write only $\hat{\mathcal{I}}$.

The next Lemma is a well known result giving analytical representations to the LT of the PPP Interference fields \cite{BacStoGeoWirNet2009}.

\begin{Lem}\label{LTexpr}
The LTs of $\hat{\mathcal{I}}^{(1)}(\rho)$ and $\hat{\mathcal{I}}^{(2)}(\rho)$, denoted by $\mathcal{L}_{\hat{\mathcal{I}}^{(1)}}(s;\rho)$ and $\mathcal{L}_{\hat{\mathcal{I}}^{(2)}}(s;\rho)$, are given by   
\begin{subequations}\label{LPempty}
\begin{align}
& \mathcal{L}_{\hat{\mathcal{I}}^{(1)}} (s;\rho) = e^{-\lambda 2 \pi (1-\delta)\int^\infty_\rho \left(1-\mathcal{L}_f(s;r) \right)r dr}, \label{LPemptya}\\
& \mathcal{L}_{\hat{\mathcal{I}}^{(2)}} (s;\rho) = e^{- \pi \lambda \delta\int^\infty_{\rho}\left(1-\mathcal{L}_g(s;r,\rho)\right)r dr} \label{LPemptyb}.
\end{align}
\end{subequations}
\end{Lem}
The Lemma uses the Poisson properties of $\hat{\Phi}^{(1)}$ and $\hat{\Phi}^{(2)}$. The expressions given in equations \eqref{LPempty} are the tools which allow us to make an entire analysis of the coverage probability.

As an example, if we replace equation \eqref{LPsingles} in equation \eqref{LPemptya}, for $\rho=0$ we get the analytical representation \cite{AndATract2011}
\begin{equation}\label{LTSinglesEx1}
\begin{split}
\mathcal{L}_{\hat{\mathcal{I}}^{(1)}}(s) &  = e^{-\frac{\lambda (1-\delta) 2\pi^2 (s p)^{2/\beta}}{\beta} csc\left(\frac{2 \pi}{\beta}\right)}, \\
\end{split}
\end{equation} 
where $csc(z)$ is the cosecant function. In the same fashion, it is possible to obtain expressions for $\mathcal{L}_{\hat{\mathcal{I}}^{(1)}} (s;\rho)$ and $\mathcal{L}_{\hat{\mathcal{I}}^{(2)}} (s;\rho)$. 

\section{Coverage probability}\label{secV}

We can now make use of the PPP superposition model for the node positions to evaluate the performance of the different cooperation (or coordination) types proposed above. From now on, we use the notation $\tilde{f}(r)$ and $\tilde{g}(r,z)$ for the beneficial signal from a single BS and a pair, to differentiate  from the interference signals $f(r)$ and $g(r,z)$.

\subsection{Fixed single transmitter}

Let us suppose that there is one BS serving the typical user, whose distance to the origin is fixed and known $r_0>0$. Moreover, it serves the typical user independently of the atoms from $\hat{\Phi}^{(1)}$ and $\hat{\Phi}^{(2)}$. Then the signal emitted to the typical user is $\tilde{f}(r_0)$, and the Signal-to-Interference-plus-Noise-Ratio (SINR) at the typical user is defined by 
\begin{equation}
\mathrm{SINR}:=\frac{\tilde{f}(r_0)}{\sigma^2+\hat{\mathcal{I}}}, 
\end{equation} 
where $\sigma^2$ is the additive Gaussian noise power at the receiver and $\hat{\mathcal{I}}$ is the total interference power (see equation \eqref{totalInt}).  

\begin{Pro}
Suppose $\tilde{f}$ as in \eqref{signalsin}. Then, the success probability is given by the expression 
\begin{equation}\label{SINRProbEF}
\mathbb{P}\left( \mathrm{SINR} >T \right) = e^{-\frac{T \sigma^2 r^\beta_0}{p} } \mathcal{L}_{\hat{\mathcal{I}}^{(1)}}\left(\frac{T r^\beta_0 }{p}\right)\mathcal{L}_{\hat{\mathcal{I}}^{(2)}}\left(\frac{T r^\beta_0}{p}\right).
\end{equation}
\end{Pro}
The last proposition allows us to evaluate the SINR directly with the help of equations $\eqref{LPemptya}$ and $\eqref{LPemptyb}$ for $\rho=0$.

\subsection{Closest transmitter from $\hat{\Phi}^{(1)}$ or $\hat{\Phi}^{(2)}$ (and his daughter)} \label{closestClusterAn}

We consider that the typical user is connected to the BS at $R_1$ (see subsection \ref{nearestN}), or to the cooperating cluster (parent,daughter) at $(R_2,Z_2)$. The previous association depends on which one of them is closer to the typical user. If $R_1<\min\{R_2,Z_2\}$, the single BS at $R_1$ serves the typical user, and it emitts the signal $\tilde{f}(R_1)$. In the opposite case, if $R_2\leq R_1$ or if $Z_2 \leq R_1$, the cooperating pair at $(R_2,Z_2)$ serves the user, and it emitts the signal $\tilde{g}(R_2,Z_2)$. All the BSs not serving the typical user generate interference. Thus, 
\begin{equation}\label{SINR}
\mathrm{SINR} := 
\begin{cases}
\frac{\tilde{f}\left( R_1 \right)}{\sigma^2+\hat{\mathcal{I}}(R_1,R_1)} \ \ ; \ \ R_1<\min\{R_2,Z_2\}, \\
\frac{\tilde{g}\left(R_2,Z_2 \right)}{\sigma^2+\hat{\mathcal{I}}(R_2,R_2)} \ \ ; \ \ R_2<\min\{R_1,Z_2\},\\ 
\frac{\tilde{g}\left(R_2,Z_2 \right)}{\sigma^2+\hat{\mathcal{I}}(Z_2,R_2)} \ \ ; \ \ Z_2<\min\{R_1,R_2\}.\\ 
\end{cases}
\end{equation}
After conditioning with respect to $R_1$, $R_2$ and $Z_2$, as a result of $\tilde{f}(R_1)$, $\hat{\mathcal{I}}(R_1,R_1)$ being independent of $(R_2,Z_2)$, and $\tilde{g}(R_2,Z_2)$, $\hat{\mathcal{I}}(R_2,R_2)$, $\hat{\mathcal{I}}(Z_2,R_2)$ being independent of $R_1$, we get
\begin{equation}\label{CP2}
\begin{split}
& \mathbb{P} (\mathrm{SINR} > T )\\ 
& = \mathbb{E}\left[ \mathbb{P}\Big( \scalebox{1} { $\frac{\tilde{f}\left( R_1 \right)}{\sigma^2+\hat{\mathcal{I}}(R_1,R_1)}>T$ } \Big| \scalebox{0.8}{$R_1$} \Big)\mathbf{1}_{ \{ \scalebox{0.7}{$R_1<\min\{R_2,Z_2\}$} \}} \right] \\
& +\mathbb{E}\left[\mathbb{P}\Big( \scalebox{1}{ $\frac{\tilde{g}\left(R_2,Z_2 \right)}{\sigma^2+\hat{\mathcal{I}}(R_2,R_2)}>T$} \Big| \scalebox{0.8}{ $R_2,Z_2$}\Big)\mathbf{1}_{ \{ \scalebox{0.7}{ $R_2<\min\{R_1,Z_2$} \}}\right]\\
& +\mathbb{E}\left[\mathbb{P}\Big( \scalebox{1}{ $\frac{\tilde{g}\left(R_2,Z_2 \right)}{\sigma^2+\hat{\mathcal{I}}(Z_2,R_2)}>T$} \Big| \scalebox{0.8}{ $R_2,Z_2$} \Big)\mathbf{1}_{ \{ \scalebox{0.7}{ $Z_2<\min\{R_1,R_2\}$} \}}\right].\\
\end{split}
\end{equation}
To evaluate the preceding equation, we need the distribution of the random vector $(R_1,R_2,Z_2)$, which was given in Section \ref{secII}. Thus, it is only left to find expressions for the conditional probabilities of equation \eqref{CP2}. For a given $r>0$, because of the independence of $R_1$ from $\hat{\mathcal{I}}(R_1,R_1)$, 
\begin{align*}
\mathbb{P} \Big( \scalebox{1} { $\frac{\tilde{f}\left( R_1 \right)}{\sigma^2+\hat{\mathcal{I}}(R_1,R_1)}>T$ } \Big| \scalebox{0.8}{$R_1=r$} \Big) = 
\mathbb{P} \left(\tilde{f} \left( r \right)>T\left( \sigma^2+\hat{\mathcal{I}}(r,r) \right)\right). \nonumber 
\end{align*}
Consider $\tilde{f}(r)$ as in \eqref{signalsin}, then it follows an exponential distribution with parameter $\frac{r^\beta}{p}$. Since $\hat{\mathcal{I}}(r,r)$ is independent of $\tilde{f}(r)$,
\begin{align}\label{Fun1}
& \mathbb{P} \left(\tilde{f}(r)>T\left( \sigma^2+\hat{\mathcal{I}}(r,r) \right)\right) \nonumber \\
& = \mathbb{E}\Big[\mathbb{P}\Big(\tilde{f}(r)>T\left( \sigma^2+\hat{\mathcal{I}}(r,r) \right)\Big| \hat{\mathcal{I}}(r,r) \Big)\Big] \nonumber \\
& = e^{\frac{- T \left(r\right)^\beta}{p} \sigma^2 } \mathcal{L}_{\hat{\mathcal{I}}^{(1)}}\Bigg(T\frac{r^\beta}{p};r \Bigg) \mathcal{L}_{\hat{\mathcal{I}}^{(2)}}\Bigg(T\frac{r^\beta}{p};r \Bigg), 
\end{align}
where the deterministic functions $\mathcal{L}_{\hat{\mathcal{I}}^{(1)}}(s;\rho)$ and $\mathcal{L}_{\hat{\mathcal{I}}^{(2)}}(s;\rho)$ are given by \eqref{LPempty}. In the same fashion, for 
$r>0$ and $z>0$,
\begin{align}
\mathbb{P} & \Big( \scalebox{1} { $\frac{\tilde{g} \left(R_2,Z_2 \right)}{\sigma^2+\hat{\mathcal{I}}(R_2,R_2)}>T$ } \Big| \scalebox{0.8}{$R_2=r,Z_2=z$} \Big) \\ & = 
\mathbb{P} \left(\tilde{g} \left(r,z\right)>T\left( \sigma^2+\hat{\mathcal{I}}(r,r) \right)\right). \nonumber 
\end{align}
For the cooperation signal $\tilde{g}(r,z)$, we use the general expression in \eqref{TAIL}. Then,
\begin{align}\label{Fun2}
& \mathbb{P}\Big( \tilde{g}(r,z)  >T\left( \sigma^2+\hat{\mathcal{I}}(r,r)\right) \Big) = \nonumber \\  
& \sum^n_{i=1}c_i \big(r,z\big) e^{-T d_i(r,z)\sigma^2} \mathcal{L}_{\hat{\mathcal{I}}^{(1)}}\big(Td_i(r,z);r\big)\mathcal{L}_{\hat{\mathcal{I}}^{(2)}}\big(Td_i(r,z);r\big). 
\end{align}
Similarly for $\mathbb{P} \Big( \scalebox{1} { $\frac{g \left(R_2,Z_2 \right)}{\sigma^2+\hat{\mathcal{I}}(Z_2,R_2)}>T$ } \Big| \scalebox{0.8}{$R_2,Z_2$} \Big)$.

\section{Numerical Evaluation} \label{secVI}
We evaluate only the noiseless scenario $\mathbb{P}(\mathrm{SIR}>T)$ (with $\sigma^2=0$). We compare the $\mathrm{SIR}$ coverage performance from the NN and the superposition model with the model without cooperation \cite{AndATract2011}. The density of the BSs is $\lambda=0.25$ [$km^2$], which corresponds to an average closest distance of $(2\sqrt{\lambda})^{-1}=1$ [km] between stations. The power is $p=1$ [$Watt$]. We consider both cases $(a)$ with fixed transmitter, and $(b)$ where the association is done with the (almost) closest cluster, as in (\ref{SINR}). In this second case,  for the NN model, the user-cluster association is done differently than in the superposition model, as follows. The typical user is served by the closest BS of the original point process $\Phi$, and by its mutually nearest neighbour, if one exists. The cooperative signals are those proposed in \eqref{coopFunction}.

\subsection{Closeness of the approximation by the PPP superposition}

\begin{figure*}[ht!]
\centering
\subfigure{\includegraphics[width=0.35\textwidth]{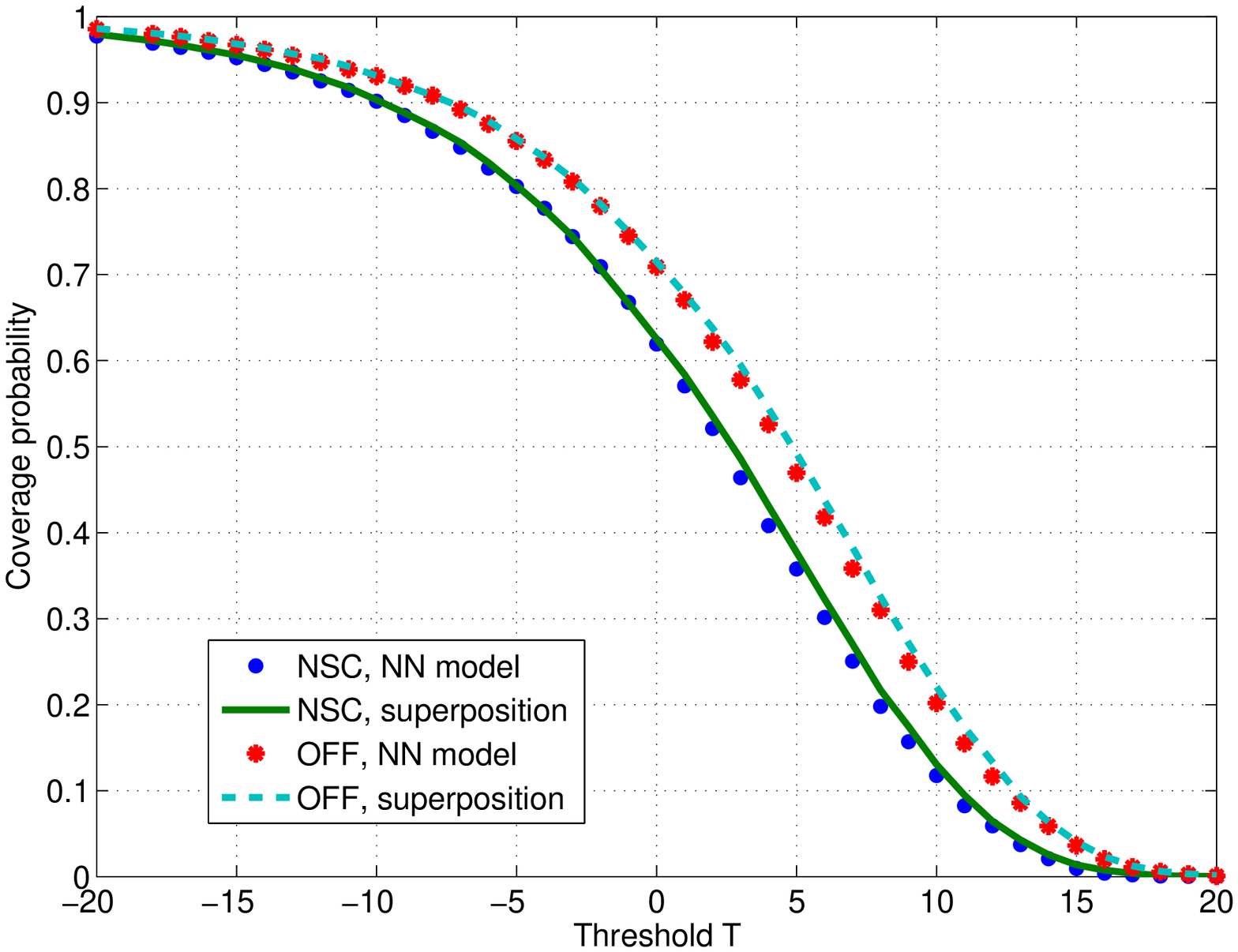}}
\subfigure{\includegraphics[width=0.35\textwidth]{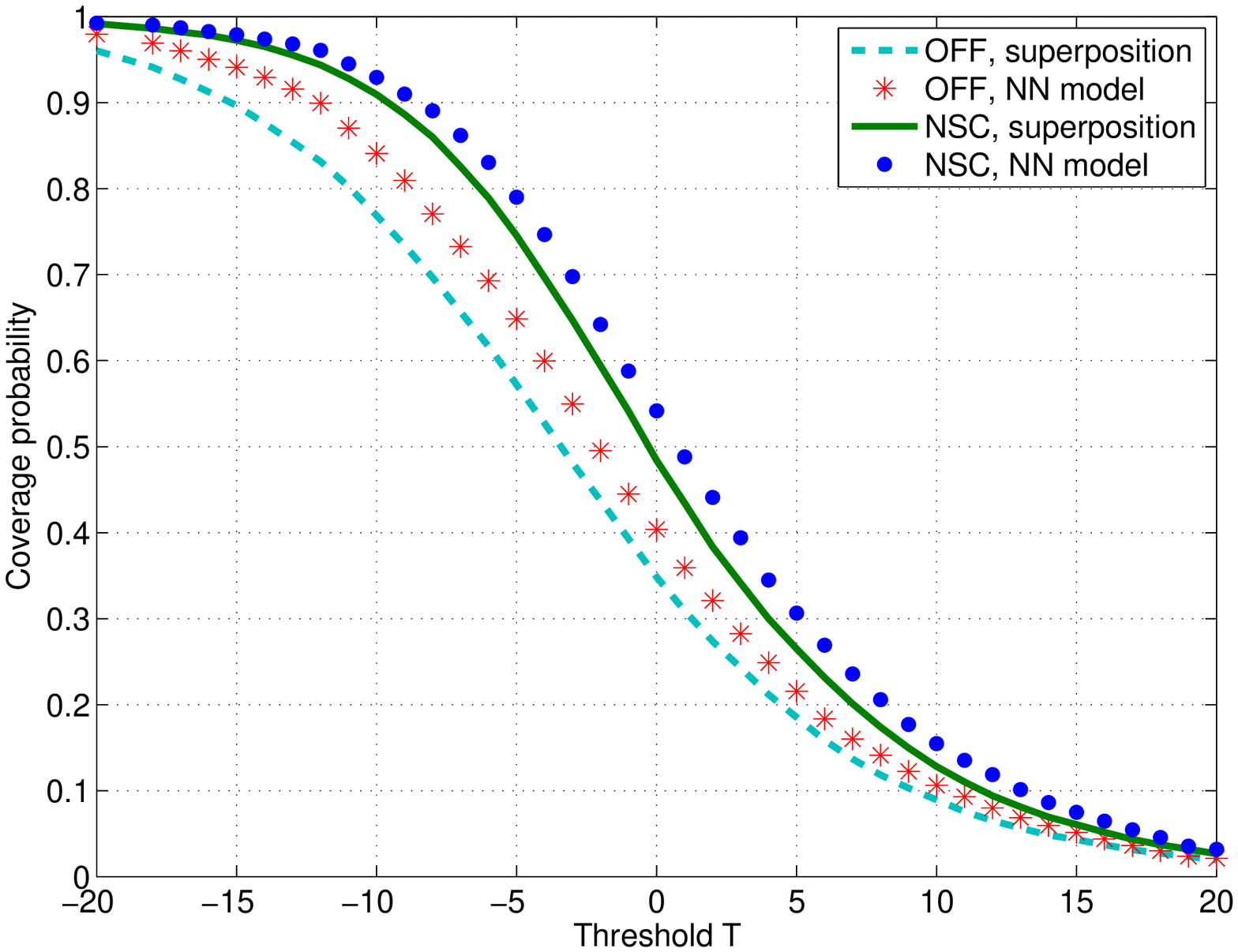}}
\caption{Closeness of the approximation between the superposition and the NN model, $\beta=3$. (a) \textit{Left:} Fixed, (b) \textit{Right:} Closest transmitter.} \label{NNVsSPCom}
\end{figure*}

We first compare in Fig. \ref{NNVsSPCom} the coverage probabilities, over the threshold $T$, for the NN model and the superposition model, in both association cases. As we can see, the curves are very close in both cases. For the \textit{"closest" transmission cluster} the difference is more evident, because on the one hand the superposition model does not take into account the repulsion between clusters (singles or pairs), and on the other hand the association of a cluster to the user as done in \eqref{SINR} for the superposition model, sometimes misses the actual closest daughter to the origin (which is not necessarily the one at $Z_2$). This never happens the way we choose the closest cluster in the NN model. Hence, the approximative model underestimates the coverage benefits in the closest cluster association. 

%

\subsection{Cooperation gains}

\begin{figure*}[t!]
\centering
\subfigure{\includegraphics[width=0.35\textwidth]{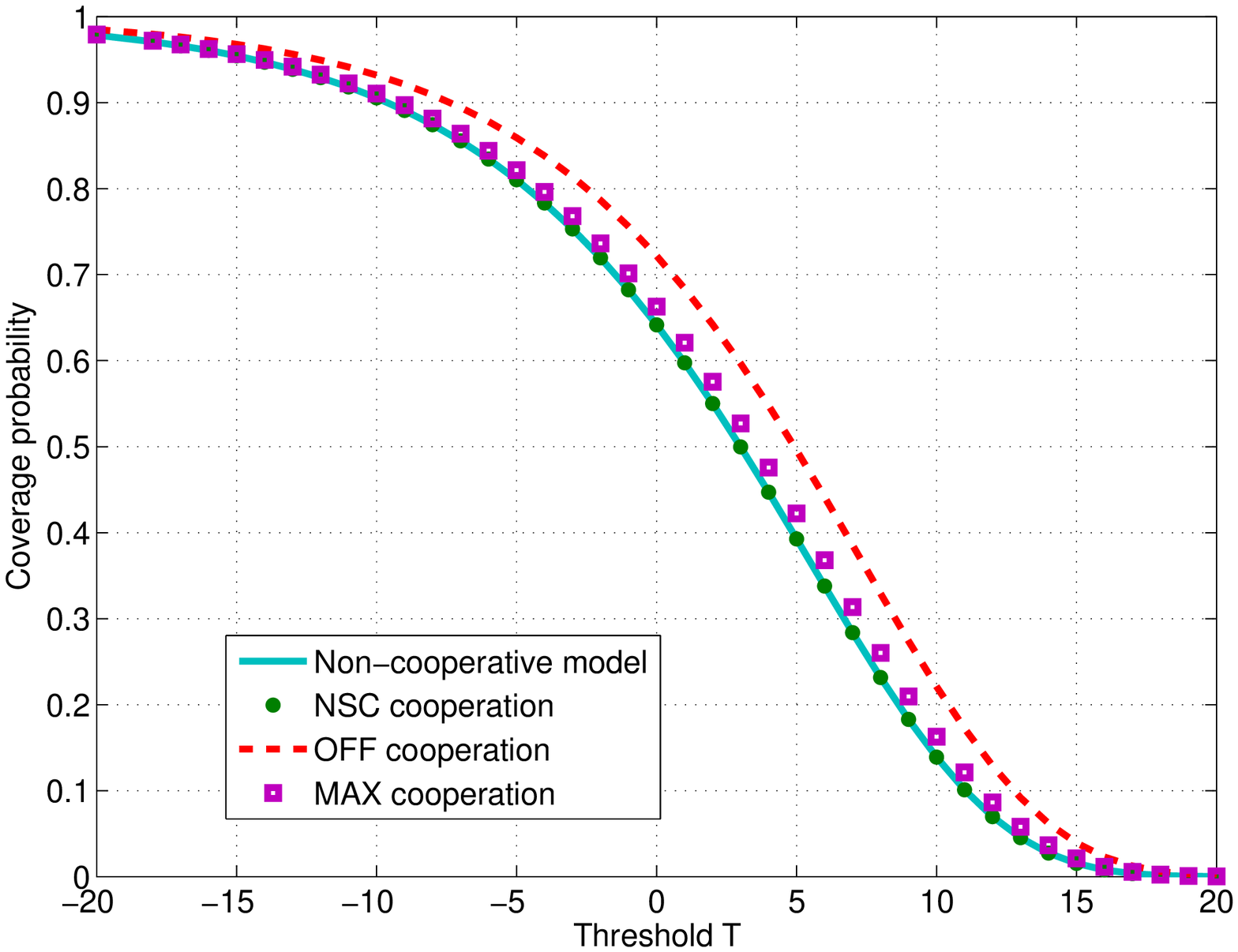}}
\subfigure{\includegraphics[width=0.35\textwidth]{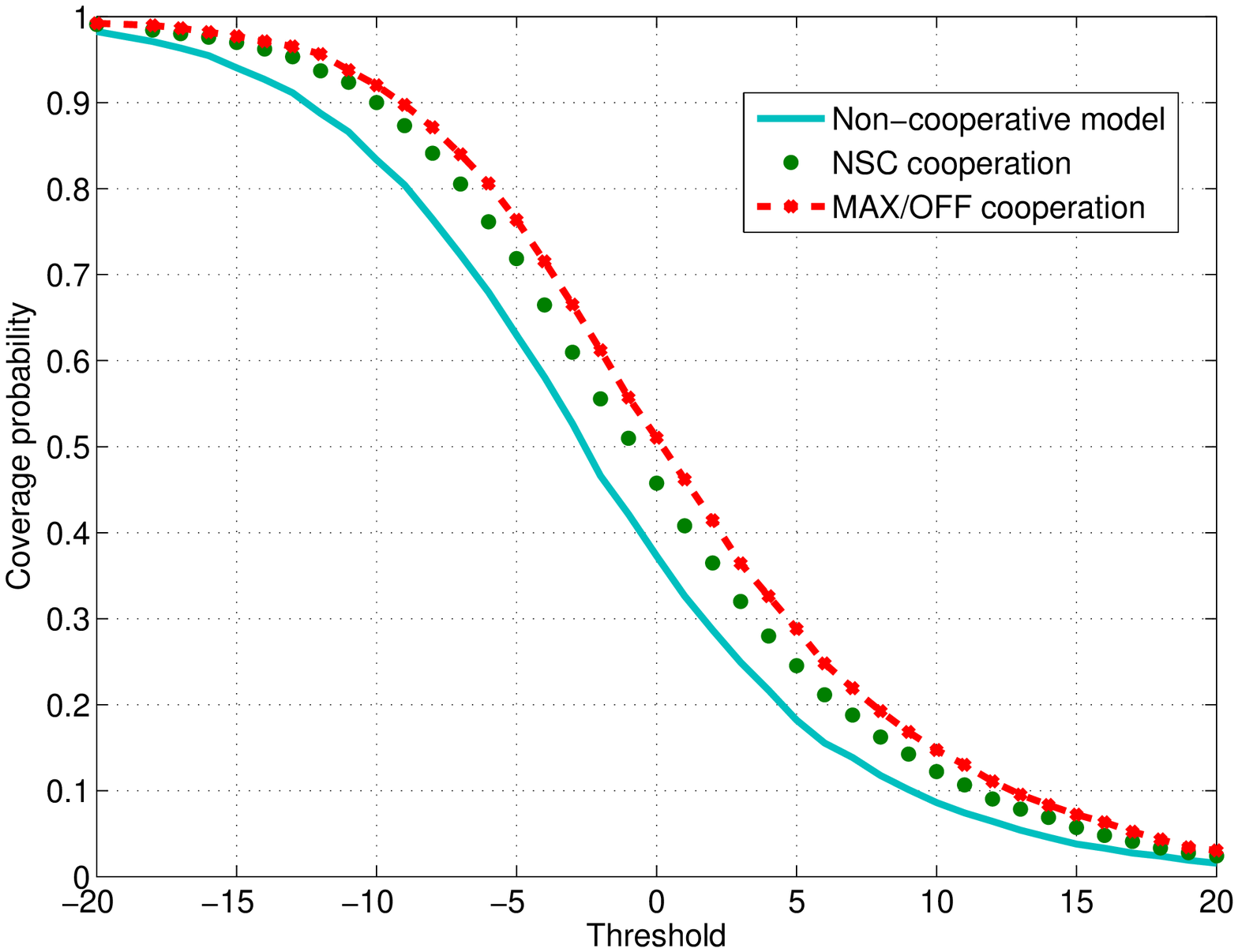}}
\caption{Coverage gains from NN cooperation compared to no cooperation, $\beta=3$. (a) \textit{Left:} Fixed, (b) \textit{Right:} Closest transmitter.} \label{CoopVsNoCoppNN}
\end{figure*}

The possible coverage gains, compared to the non-cooperative network, in the case of association with a fixed transmitter, are shown in Fig. \ref{CoopVsNoCoppNN}(a). As a first remark, for the fixed association, the \textbf{[NSC]} case for the NN model and the non-cooperative model are practically the same. The coverage probability in the \textbf{[MAX]} case is close to the coverage probability in the \textbf{[NSC]} case. This suggests that the strongest signal in each cooperating pair influences interference the most. For the \textbf{[OFF]} case there is a $10\%$ benefit compared to the non-cooperative case, in the largest part of the domain in $T$.

The gains are also evaluated in the case of association with the closest cluster. For the SINR, let us call \textbf{[MAX/OFF]} the case where the closest cluster emits a signal to the typical user according to \textbf{[MAX]}, i.e. only the max signal is sent, while the pairs generate interference according to \textbf{[OFF]}. The idea is that when all network pairs choose \textbf{[MAX]} cooperation for their own users, this choice of one-station-out-of-two is random for the typical user point of view. This \textbf{[MAX/OFF]} case shows a $15\%$ absolute gain from the non cooperative case, which is around $9\%$ for the \textbf{[NSC]} (see Fig. \ref{CoopVsNoCoppNN}(b)). This gain is almost equal with the dynamic clustering in \cite{BacAStoGeo2015}.

\section{Conclusions}\label{secVII}
We have proposed a general methodology to evaluate coverage from clusters of mutually nearest neighbours, which involves only $62\%$ of the BSs. Such cooperation between BSs gives considerable coverage benefits, even for clusters of size at most two. Further analysis on higher order ($>2$) cooperation clusters, as well as different cooperation schemes (e.g. MIMO), could reveal even greater potentials. 


\bibliographystyle{unsrt}



\section*{Appendix}

Let $R$ be a Rayleigh random variable, with scale parameter $a>0$. If we denote by $F_R$ and $f_R$ its CDF and PDF functions, respectively, then, for every $r>0$,  
\begin{equation}\label{Rayliegh}
\begin{split}
F_R(r)=1-e^{-\frac{r^2}{2 a^2}}, \ \ f_R(r)=\frac{r}{a^2}e^{-\frac{r^2}{2a^2}}.\\
\end{split}
\end{equation}

\begin{figure}[htbp]
\centering
\subfigure[]{\includegraphics[width=0.25\textwidth]{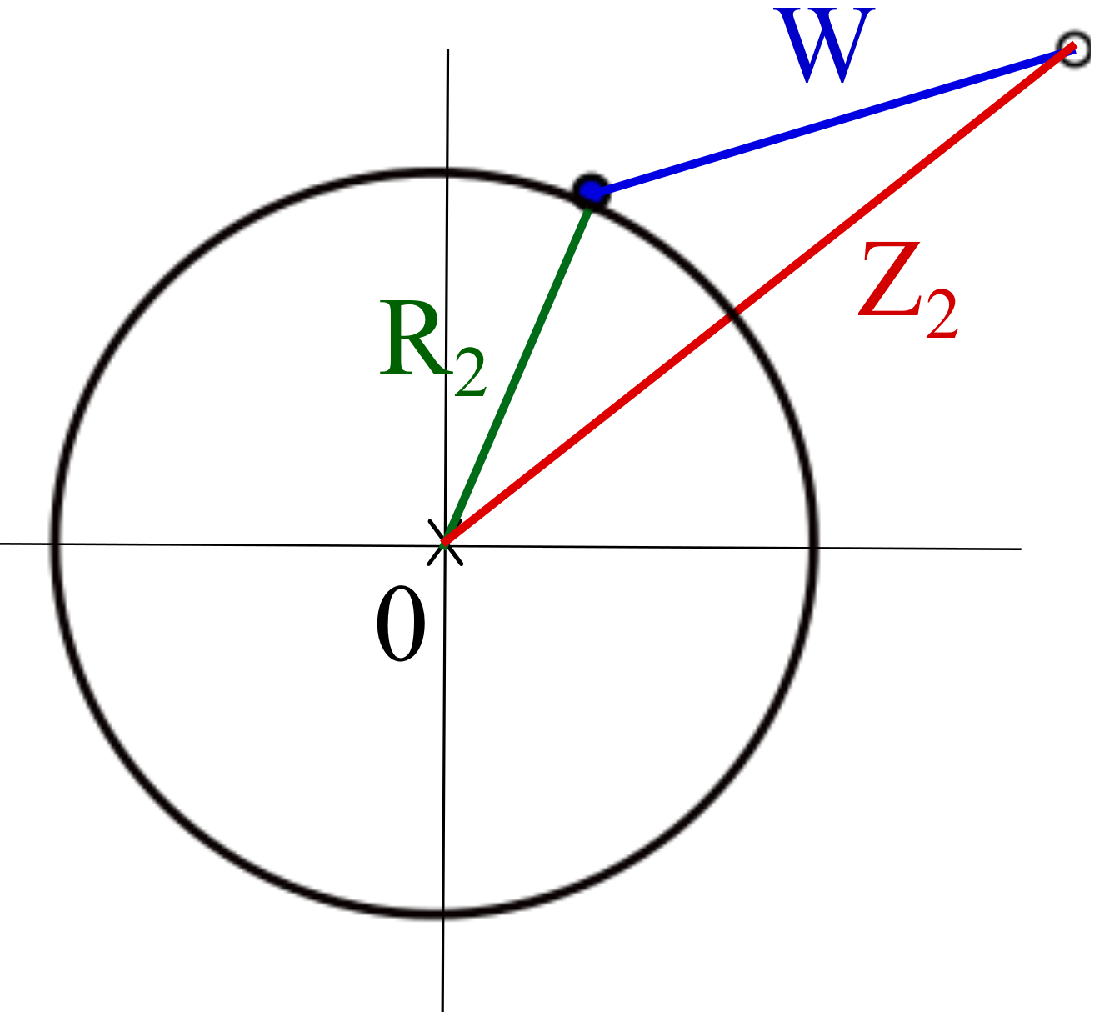}}
\caption{}\label{prueba}
\end{figure}

Denote by $A$ and $B$ the Cartesian coordinates of the nearest parent to the typical user and his daughter, respectively, and also denote by $(R_2,\Theta)$ and $(Z_2,\Psi)$ their respective polar coordinates. Define $C:=A-B$ and denote its polar coordinates by $(W,\Omega)$ (see Figure \ref{prueba}). 

As stated in Section \ref{secII}, the random variables $R_2$ and $W$ are Rayleigh distributed, with scale parameters $\zeta$ and $\alpha$, respectively. Moreover, the random angles $\Theta$, $\Psi$ and $\Omega$ are considered uniformly distributed over $[0,2\pi)$, to preserve the isotropy in the PPP case. Also, the random variables $R_2$, $\Theta$, $W$, and $\Omega$ are independent between them, as in the PPP case. 

Our first goal is to find the joint distribution of the random vector $(R_2,Z_2)$ and, as a consequence, find also the distribution of $Z_2$. To do so, let us remark that the cartesian coordinates of a point around a center, wich has rayleigh radial distance from the origin and uniform angle, are distibuted as an independent Gaussian vector \cite[pp. 276, Ex. 7b]{AfirstCourseRoss}. Hence, there exist independent random variables $A_x,A_y,C_x,C_y$, where $A_x,A_y$ are Normal distributed, with parameter $(0,\zeta^2)$, and $C_x,C_y$ are also Normal distributed, with parameters $(0,\alpha^2)$, and such that 
\begin{equation*}
\begin{split}
(A_x,A_y) = (R_2cos\Theta,R_2sin\Theta), \\
(C_x,C_y) = (Wcos\Omega,Wsin\Omega).
\end{split}
\end{equation*}

\subsection{Proof of Lemma \ref{denR2ZR2}.} 
By definition, 
\begin{equation*}
\begin{split}
A_x & = R_2 cos\Theta, \ \ \ \  C_x = R_2cos\Theta-Z_2cos\Psi, \\
A_y & = R_ 2 sin\Theta, \ \ \ \ C_y = R_2sin\Theta-Z_2 sin\Psi.
\end{split}
\end{equation*}
The absolute value of the Jacobian of the above transformation is $R_2 Z_ 2$. Let us denote by  $f_{A_x,A_y,C_x,C_y}$ the joint PDF of $(A_x,A_y,C_x,C_y)$, if $f_{R,\Theta,Z,\Psi}$ denotes the joint PDF  of $(R_2,\Theta,Z_2,\Psi)$, then, by the change of variable Theorem \cite[pp. 274]{AfirstCourseRoss},  
\begin{equation*}
\begin{split}
f & _{R,\Theta,Z,\Psi} (r,\theta,z,\psi) \\
& = f_{A_x,A_y,C_x,C_y}(rcos\theta,rsin\theta,rcos\theta-zcos\psi,rsin\theta-zsin\psi)rz \\
& \stackrel{(a)}{=} \frac{rz}{(2\pi\alpha \zeta)^2}e^{-\left( \frac{ r^2cos^2\theta}{2\zeta^2}+\frac{ r^2sin^2\theta}{2\zeta^2}+\frac{ (rcos\theta-zcos\psi)^2}{2\alpha^2}+\frac{ (rsin\theta-zsin\psi)^2}{2\alpha^2} \right)} \\
& \stackrel{(b)}{=} \frac{rz}{(2\pi\alpha \zeta)^2}e^{-\left( \frac{r^2}{2} \left(\frac{1}{\alpha^2}+\frac{1}{\zeta^2}\right)
+\frac{ z^2}{2\alpha^2}-\frac{ rzcos(\theta-\psi)}{\alpha^2} \right)},
\end{split}
\end{equation*}
where $(a)$ comes from the formula of the distribution of independent Gaussian random variables, and $(b)$ follows from the trigonometric identities $cos^2\theta+sin^2\theta =1$ and $cos\theta cos\psi+sin\theta sin\psi =cos(\theta-\psi)$. To obtain the joint PDF of $(R_2,Z_2)$, denoted by $f_{R_2,Z_2}$, we integrate the previous expression over $[0,2\pi)\times [0,2\pi)$, with respect to the variables $\theta$ and $\psi$, 
\begin{equation*}
\begin{split}
f & _{R_2,Z_2} (r,z) \\
& \stackrel{(c)}{=} \frac{rz}{( \alpha \zeta)^2}e^{-\left( \frac{r^2}{2} \left(\frac{1}{\alpha^2}+\frac{1}{\zeta^2}\right)+\frac{z^2}{2\alpha^2} \right) } \frac{1}{2\pi}\int^{2\pi}_0  e^{\frac{ rzcos w}{\alpha^2} }dw \\
& \stackrel{(d)}{=} \frac{rz}{(\alpha \zeta)^2}e^{-\left( \frac{r^2}{2} \left(\frac{1}{\alpha^2}+\frac{1}{\zeta^2}\right)+\frac{z^2}{2\alpha^2} \right) } I_0\left(\frac{rz}{\alpha^2} \right),
\end{split}
\end{equation*}
where $(c)$ comes from the change of variable $w=\theta-\psi$ and $(d)$ follows because the integral representation $I_0(x)=\frac{1}{2\pi}\int^{2\pi}_0  e^{x cosw }dw$ \cite{ModBessEric}. Let us denote by $f_{Z_2}$ the PDF of the random variable $Z_2$ and by $\eta=\left(\frac{1}{\alpha^2}+\frac{1}{\zeta^2}\right)$. To obtain $f_{Z_2}$, we integrate over $[0,\infty)$ with respect to the variable $r$ the preceding equation
\begin{align*}
f_{Z_2}(z) & \stackrel{(e)}{=} \int^\infty_0 \frac{rz}{(\alpha\zeta)^2}e^{-\left(\frac{r^2}{2} \eta+\frac{z^2}{2\alpha^2}\right)}\sum^\infty_{n=0}\frac{(1/4)^n}{(n!)^2}\left( \frac{rz}{\alpha^2} \right)^{2n} dr \\
& =  \frac{z}{(\alpha\zeta)^2}e^{-\frac{z^2}{2\alpha^2}} \sum^\infty_{n=0} \frac{(1/4)^n}{(n!)^2}\left( \frac{z^2}{\alpha^4} \right)^{n}  \int^\infty_0 r^{2n} e^{-\frac{r^2}{2}  \eta} r dr \\
& \stackrel{(f)}{=} \frac{z}{(\alpha\zeta)^2\eta}e^{-\frac{z^2}{2\alpha^2}} \sum^\infty_{n=0} \frac{\left(\frac{z^2}{2 \alpha^4 \eta }\right)^n}{n!} \\
& = \frac{z}{(\alpha\zeta)^2\eta} e^{-\frac{z^2}{2\alpha^2} }e^{\frac{z^2}{2 \alpha^4\eta} } \\
& \stackrel{(g)}{=}  \frac{z}{\alpha^2+\zeta^2} e^{-\frac{z^2}{2(\alpha^2+\zeta^2) } }, \\
\end{align*}
where $(e)$ comes from the series representation $I_0(x)=\sum^\infty_{n=0}\frac{(1/4)^n}{(n!)^2}x^{2n}$ \cite{ModBessEric}, while $(f)$ follows after the formula
\begin{equation*}
\int^\infty_0 r^{2n} e^{-\frac{r^2}{2}\eta} r dr=\frac{2^n}{\eta^{n+1}}n!,
\end{equation*}
and $(g)$ after soma algebraic manipulations and from the definition of $\eta$.

\subsection{Initial calculations for the closest cluster association}

In this subsection, we provide the calculations to obtain the coverage probability in equation \eqref{CP2}  from the definition of the SINR in equation \eqref{SINR}. We begin with the analysis for the first term with the non-cooperative signal
 
\begin{equation*}
\begin{split}
\mathbb{E} &\Bigg[  \mathbf{1}_{\Big\{\frac{\tilde{f}\left( R_1 \right)}{\sigma^2+\hat{\mathcal{I}}(R_1,R_1)}>T \Big\} }  \mathbf{1}_{\{R_1<\min{\{R_2,Z_2\} } \} }   \Bigg]  \\
&  = \mathbb{E}  \Bigg[ \mathbb{E}\Bigg[  \mathbf{1}_{\Big\{\frac{\tilde{f}\left( R_1 \right)}{\sigma^2+\hat{\mathcal{I}}(R_1,R_1)}>T \Big\} }  \mathbf{1}_{\{R_1<\min{\{R_2,Z_2\} } \} }\Big | R_1,R_2,Z_2 \Bigg] \Bigg]  \\
& \stackrel{(a)}{=} \mathbb{E} \Bigg[ \mathbf{1}_{\{R_1<\min{\{R_2,Z_2\} } \} } \mathbb{E}\Bigg[  \mathbf{1}_{\Big\{\frac{\tilde{f}\left( R_1 \right)}{\sigma^2+\hat{\mathcal{I}}(R_1,R_1)}>T \Big\} } \Big | R_1,R_2,Z_2 \Bigg] \Bigg]  \\
& \stackrel{(b)}{=} \mathbb{E} \Bigg[ \mathbf{1}_{\{R_1<\min{\{R_2,Z_2\} } \} } \mathbb{E}\Bigg[  \mathbf{1}_{\Big\{\frac{\tilde{f}\left( R_1 \right)}{\sigma^2+\hat{\mathcal{I}}(R_1,R_1)}>T \Big\} } \Big | R_1 \Bigg] \Bigg]  \\
& \stackrel{(c)}{=} \mathbb{E} \Bigg[ \mathbf{1}_{\{R_1<\min{\{R_2,Z_2\} } \} }  \mathbb{P}\Bigg( \frac{\tilde{f}\left( R_1 \right)}{\sigma^2+\hat{\mathcal{I}}(R_1,R_1)}>T  \Big | R_1 \Bigg) \Bigg],  
\end{split}
\end{equation*}
where $(a)$ comes from properties of the conditional expectation, $(b)$ follows because the random variable $\frac{\tilde{f}\left( R_1 \right)}{\sigma^2+\hat{\mathcal{I}}(R_1,R_1)}>T$ is independent of $R_2$ and $Z_2$, and $(c)$ because $\mathbb{E}[\textbf{1}_B|Y]=\mathbb{P}(B|Y)$.

After a similar analysis for the two terms with the cooperative signal 

\begin{equation*}
\begin{split}
\mathbb{E} &\Bigg[  \mathbf{1}_{\Big\{\frac{\tilde{g}\left( R_2,Z_2 \right)}{\sigma^2+\hat{\mathcal{I}}(R_2,R_2)}>T \Big\} }  \mathbf{1}_{\{R_2<\min{\{R_1,Z_2\} } \} }   \Bigg]  \\
& = \mathbb{E} \Bigg[ \mathbf{1}_{\{R_2<\min{\{R_1,Z_2\} } \} }  \mathbb{P}\Bigg( \frac{\tilde{g}\left( R_2,Z_2 \right)}{\sigma^2+\hat{\mathcal{I}}(R_2,R_2)}>T  \Big | R_1 \Bigg) \Bigg], \\
\mathbb{E} &\Bigg[  \mathbf{1}_{\Big\{\frac{\tilde{g}\left( R_2,Z_2 \right)}{\sigma^2+\hat{\mathcal{I}}(Z_2,R_2)}>T \Big\} }  \mathbf{1}_{\{Z_2<\min{\{R_1,R_2\} } \} }   \Bigg]  \\
& = \mathbb{E} \Bigg[ \mathbf{1}_{\{Z_2<\min{\{R_1,R_2\} } \} }  \mathbb{P}\Bigg( \frac{\tilde{g}\left( R_2,Z_2 \right)}{\sigma^2+\hat{\mathcal{I}}(Z_2,R_2)}>T  \Big | R_1 \Bigg) \Bigg],  
\end{split}
\end{equation*}

\subsection{Further calculations for the closest cluster association}

Denote by 
\begin{equation*}
\begin{split}
& G(r) := \mathbb{P}\Bigg( \frac{\tilde{f}\left( R_1 \right)}{\sigma^2+\hat{\mathcal{I}}(R_1,R_1)}>T \Big| R_1=r \Bigg) \\
& H(r,z) := \mathbb{P}\Bigg( \frac{\tilde{g}\left(R_2,Z_2 \right)}{\sigma^2+\hat{\mathcal{I}}(R_2,R_2)}>T \Big|R_2=r,Z_2=z\Bigg) \\
& K(r,z) := \mathbb{P}\Bigg( \frac{\tilde{g}\left(R_2,Z_2 \right)}{\sigma^2+\hat{\mathcal{I}}(Z_2,R_2)}>T \Big| R_2=r,Z_2=z \Bigg).
\end{split}
\end{equation*}
Remark that at the end of section \ref{secV} we obtained the expressions for the deterministic functions $G(r)$, $H(r,z)$ and $K(r,z)$ (see equations \eqref{Fun1} and \eqref{Fun2}). To complete the analysis in subsection \ref{closestClusterAn}, we need to find the coverage probability expressed in equation \eqref{CP2}, thus, we need expressions for   
\begin{equation*}
\begin{split}
\mathbb{E} & \left[G(R_1)\textbf{1}_{\{R_1<\min\{R_2,Z_2\}\}}\right], \\
\mathbb{E} & \left[H(R_2,Z_2)\textbf{1}_{\{R_2<\min\{R_1,Z_2\}\}}\right], \\
\mathbb{E} & \left[K(R_2,Z_2)\textbf{1}_{\{Z_2<\min\{R_1,R_2\}\}}\right]. \\
\end{split}
\end{equation*}
Let us begin by the first one,
\begin{equation*}
\begin{split}
\mathbb{E} & \left[G(R_1)\textbf{1}_{\{R_1<\min{R_2,Z_2}\}}\right] \\
& \stackrel{(a)}{=} \mathbb{E}\left[ \mathbb{E}\left[ G(R_1)\textbf{1}_{\{R_1<\min{R_2,Z_2}\}}|R_1\right] \right] \\
& = \mathbb{E}\left[ G(R_1) \mathbb{E}\left[ \textbf{1}_{\{R_1<\min{R_2,Z_2}\}}|R_1\right] \right] \\
& \stackrel{(b)}{=} \mathbb{E}\left[ G(R_1) \mathbb{P}\left( \min\{R_2,Z_2\}>R_1|R_1\right) \right],
\end{split}
\end{equation*}
where $(a)$ follows by properties of the conditional expectation and $(b)$ because $\mathbb{E}[\textbf{1}_B|Y]=\mathbb{P}(B|Y)$. Thus, we only have left to find an expression for $\mathbb{P}\left( \min\{R_2,Z_2\}>R_1|R_1=r\right)$. Because $R_1$ is independent of $(R_2,Z_2)$, 
\begin{equation*}
\mathbb{P}(\min{\{R_2,Z_2\}}>R_1|R_1=r) = \mathbb{P}(\min{\{R_2,Z_2\}}>r),
\end{equation*}
and then  
\begin{equation*}
\begin{split}
\mathbb{P}(\min{\{R_2,Z_2\}}>r) = 1-F_{R_2}(r)-F_{Z_2}(r)+F_{R_2,Z_2}(r,r),
\end{split}
\end{equation*}
where $F_{R_2}$, $F_{Z_2}$, and $F_{R_2,Z_2}$ are the CDF of $R_2$, $Z_2$, and $(R_2,Z_2)$ that can be explicitly obtained after considering equations \eqref{Rayliegh} and \eqref{jointDensityFunction}.

In the same fashion,
\begin{equation*}
\begin{split}
\mathbb{E} & \left[H(R_2,Z_2)\textbf{1}_{\{R_2<\min{R_1,Z_2}\}}\right] \\
& = \mathbb{E} \left[H(R_2,Z_2)\mathbb{P}(\min\{R_1,Z_2\}>R_2|R_2,Z_2)\right], \\
\mathbb{E} & \left[ K(R_2,Z_2)\textbf{1}_{\{Z_2<\min{R_1,R_2}\}} \right] \\
& = \mathbb{E} \left[K(R_2,Z_2)\mathbb{P}(\min\{R_1,R_2\}>Z_2|R_2,Z_2)\right] \\
\end{split}
\end{equation*}
and
\begin{equation*}
\begin{split}
\mathbb{P}(\min{\{R_1,Z_2\}}>R_2|R_2=r,Z_2=z) & = (1-F_{R_1}(r))\mathbf{1}_{\{z>r\}}, \\
\mathbb{P}(\min{\{R_1,R_2\}}>Z_2|R_2=r,Z_2=z) & = (1-F_{R_1}(z))\mathbf{1}_{\{r>z\}},
\end{split}
\end{equation*}
where $F_{R_1}$ is the CDF of $R_1$ that can be explicitly obtained after considering equation \eqref{Rayliegh}.

Having done this, we can evaluate the coverage probability given by equation \eqref{CP2}. For example, to evaluate 
\begin{equation*}
\mathbb{E} \left[H(R_2,Z_2)\mathbb{P}(\min\{R_1,Z_2\}>R_2|R_2,Z_2)\right],
\end{equation*}
we use the explicit form of the function $H(r,z)$, given in equation \eqref{Fun2}, with the explicit form of 
\begin{equation*}
\mathbb{P}(\min\{R_1,Z_2\}>R_2|R_2=r,Z_2=z)=(1-F_{R_1}(r))\mathbf{1}_{\{z>r\}},
\end{equation*}
and the joint distribution of $(R_2,Z_2)$, given in Lemma \ref{denR2ZR2}.
\subsection{Validity of the numerical analysis.}

\begin{figure}[htbp]
\centering
\subfigure[$\beta=2.5$]{\includegraphics[width=0.35\textwidth]{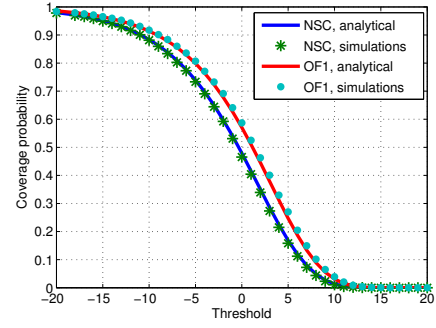}}
\subfigure[$\beta=4$]{\includegraphics[width=0.35\textwidth]{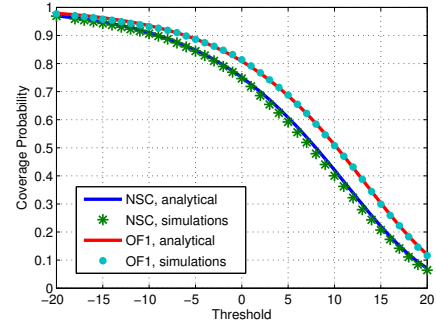}}
\caption{Validity of the analysis for the superposition model for the fixed single transmitter.}\label{AnalVsSim} 
\end{figure}

In Figure \ref{AnalVsSim} we compare the plots of the coverage probability from the numerical integration of the analytical formula in \eqref{SINRProbEF}, with simulations for the superposition model, for the fixed single transmitter case.

\begin{figure}[htbp]
\centering
\subfigure[$\beta=2.5$]{\includegraphics[width=0.35\textwidth]{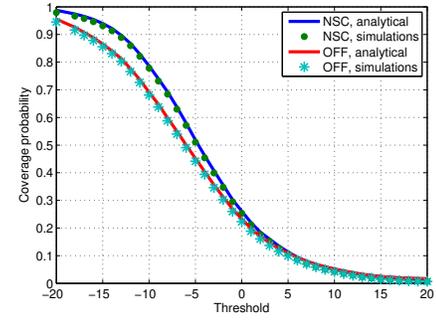}}
\subfigure[$\beta=4$]{\includegraphics[width=0.35\textwidth]{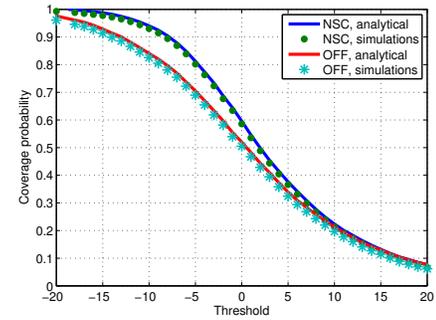}}
\caption{Validity of the analysis for the superposition model for the closest cluster association.} \label{AnalVsSim2} 
\end{figure}

In Figure \ref{AnalVsSim2} we compare the plots of the coverage probability from the numerical integration of the analytical formula in \eqref{CP2}, with simulations for the superposition model, for the closest cluster case transmitter defined in subsection \ref{closestClusterAn}.

As we can see, the simulations and the numerical analytical results fit perfectly, both for larger values of $\beta$, like $\beta=4$, and for critical ones, like $\beta=2.5$.

\end{document}